# Partition of Optical Properties Into Orbital Contributions

Sebastian P. Sitkiewicz,[a,b,c] Mauricio Rodriguez-Mayorga,[a,b,c] Josep M. Luis,[*c]
Eduard Matito[*b,d]

Nonlinear optical properties (NLOPs) play a major role in photonics, electro-optics and optoelectronics, and other fields of modern optics. The design of new NLO molecules and materials has benefited from the development of computational tools to analyze the relationship between the electronic structure of molecules and its optical response. In this paper, we present a new means to analyze the response property through the partition of NLOPs in terms of orbital contributions (PNOC). This tool can be used to obtain a local representation of the NLOPs, providing a powerful visualization aid to connect the magnitude of the optical property with some parts of the molecule. Unlike other methods to analyze NLOPs, the PNOC decomposes the optical property into orbitals of the unperturbed system, furnishing this method with the ability to assess the performance of single- and multi-determinant electronic structure methods. PNOC can be also used to design small basis sets for an accurate description of large systems, saving a substantial amount of computer time for the calculation of optical properties.

## 1 Introduction

Nonlinear optical properties (NLOPs) have been extensively studied since the discover of the second-harmonic generation exhibited by a quartz sample irradiated by a ruby laser.[1] This discovery led to applications in optical communications, photonics, and optoelectronics, such as X-ray generation, the study of correlated photon pairs or biological imaging, spurring the quest for molecules and materials with high nonlinear optical (NLO) response.[2]

The design of new NLO materials and molecules has benefited from the evolution of quantum chemistry, which has provided the framework to compute NLOPs. In addition, the development of computational tools to analyze the relationship between the electronic structure of molecules and its optical response has greatly contributed to guide the design of nonlinear materials. Perhaps the simplest of such tools is the well-known two-level approach of Chemla and Oudar,[3] which simplifies the expression of the calculation of the hyperpolarizability, unveiling the most important molecular features that are responsible for the enhancement of the second-order nonlinear response. In the same vein, the sum-over-states (SOS) technique,[4] which provides a perturbation theory approach to compute NLOPs, is currently still used as a way to assess the role of some electronic states in the magnitude of the NLOP. Additional works also considered the removal of states[5] or orbitals[6] in the SOS expression to estimate their importance.

One of the landmark papers on the analysis of NLOPs is due to Chopra and co-workers,[7] who studied orbital contributions to linear and NLOPs. Although the contributions sum up to the total NLOP value, they depend on the position of the origin of coordinates axis. Despite this unsettling downside, origin-dependent methods are widespread and the coordinate axis is usually fixed at the center of mass. Nakano et al.[8–10] defined the contributions to the $n$-order optical property as the $n$-order derivative of the density with respect to the electric field. Obviously, the sum of these quantities does not add up to the total value of the NLOP, but they are not origin dependent. The analysis has been extended to open-shells systems[11,12] using the density of effectively unpaired electrons,[13] which gives rise to pairwise contributions to the optical property.

[a] *Kimika Fakultatea, Euskal Herriko Unibertsitatea (UPV/EHU), 20080 Donostia, Euskadi, Spain*
[b] *Donostia International Physics Center (DIPC), Donostia, Euskadi, Spain*
[c] *Institut de Química Computacional i Catàlisi (IQCC) and Departament de Química, Universitat de Girona, C/ Maria Aurèlia Capmany, 69, 17003 Girona, Catalonia, Spain*
[d] *Ikerbasque Foundation for Science, 48011 Bilbao, Euskadi, Spain*

More recently, Geskin and Brédas have considered NLOP contributions based on the derivatives of the Mulliken charges with respect to the electric field,[14,15] providing origin-dependent contributions that do not add to the total values for all optical properties. The latter approach was originally proposed to study the static optical response and it was later extended to dynamic properties within the frameworks of time-dependent Hartree-Fock[16] and Kohn-Sham[17] levels. Localized orbitals have been also employed to decompose NLOPs.[17,18] Other analyses provide a pairwise decomposition of the optical property. This is the case of the Hieringer and Baerends decomposition of the first hyperpolarizability within the framework of the Kohn-Sham density functional theory[19] and the recent work of Mandado and co-workers using field-induced orbitals.[20] The NLO properties of interacting species can be also decomposed into terms of different physical origins like the exchange repulsion or the charge delocalization using approaches analogous to the ones defined in the symmetry-adapted perturbation theory (SAPT).[21–24] Finally, one should also mention the decomposition of NLOPs in the real space. Most of these analyses[25–30] are framed in the quantum theory of atoms in molecule (QTAIM),[31] although Hirshfeld-based atomic partitions have been also used to analyze the polarizability.[32]

Despite the importance of quantum chemistry in optics, the calculation of NLOPs poses some challenges for wavefunction and density functional approximations.[33–36] The assessment of approximate electronic structure methods is thus necessary and is often done using expensive wavefunction methods.[37–39] Decomposing NLOPs into orbital contributions would help identifying pitfalls of electronic structure methods such as the wrong selection of active orbital spaces or the adequate size of orbital basis sets.

In this paper, we present the partition of NLOPs into orbital contributions (PNOC), a new means to analyze the response property and estimate its local contributions. PNOC provides a convenient way to relate linear and nonlinear optical properties to the structure of the molecule, adding a new tool for the construction of molecules with potential applications in photonics and optoelectronics. In addition, PNOC provides a feature that is absent in many other tools to analyze NLOPs: a decomposition in terms of the orbitals of the unperturbed system that can be also used to assess the performance of electronic structure methods and design conveniently small basis sets to study large molecules. In the following we provide some background knowledge needed to introduce the PNOC and, afterwards, we give some examples that illustrate the potential of this tool.

## 2 Methodology

In the quantum–mechanical picture, under the perturbation of an electric field $\boldsymbol{F}$, the Hamiltonian of a molecular system $\hat{H}'$ is expressed as a sum of the unperturbed Hamiltonian $\hat{H}$ and the term that describes the interaction of the electrons ($a$) and nuclei ($A$) with the external electric field $\boldsymbol{F}$:[7,40–43]

$$\hat{H}' = \hat{H} + e\sum_a \boldsymbol{r}_a \cdot \boldsymbol{F} - \sum_A Z_A \boldsymbol{R}_A \cdot \boldsymbol{F} \quad (1)$$

In the present study, the electronic nonrelativistic Hamiltonian and electronic part of the wavefunction are considered, with static and homogeneous electric fields taken into consideration. This prevents the contribution of higher multipole moments (like quadrupole or octupole) to the total energy.[40,43] However, the present scheme could be easily generalized for such multipole moments.

The classic definition of the molecular static electric (hyper)polarizabilities comes from the Taylor expansion of the field–dependent dipole moment $\boldsymbol{\mu}(\boldsymbol{F})$:[7,40,41,43]

$$\mu_i(\boldsymbol{F}) = \mu_i(\boldsymbol{0}) + \sum_j \alpha_{ij} F_j + \frac{1}{2!}\sum_{jk}\beta_{ijk}F_jF_k + \frac{1}{3!}\sum_{jkl}\gamma_{ijkl}F_jF_kF_l + \cdots$$

where $i$, $j$, $k$ and $l$ can be any of the Cartesian components: $x$, $y$ or $z$. In the above equation, the permanent dipole moment ($\mu_i(\boldsymbol{0})$) is altered by the linear response, quantified by the polarizability ($\alpha_{ij}$), and the nonlinear response, represented by the first hyperpolarizability $\beta_{ijk}$ and second hyperpolarizability $\gamma_{ijkl}$, which are consecutive partial derivatives of $\mu_i(\boldsymbol{F})$:

$$\alpha_{ij} = \left.\frac{\partial \mu_i(\boldsymbol{F})}{\partial F_j}\right|_{\boldsymbol{F}=\boldsymbol{0}} \quad (2)$$

$$\beta_{ijk} = \left.\frac{\partial^2 \mu_i(\boldsymbol{F})}{\partial F_j \partial F_k}\right|_{\boldsymbol{F}=\boldsymbol{0}} \quad (3)$$

$$\gamma_{ijkl} = \left.\frac{\partial^3 \mu_i(\boldsymbol{F})}{\partial F_j \partial F_k \partial F_l}\right|_{\boldsymbol{F}=\boldsymbol{0}} \quad (4)$$

An equivalent description of the NLOPs can be obtained in the terms of the derivatives of the one–electron densities with respect to the external field utilizing the definition of the dipole moment $\mu_i(\boldsymbol{F})$ in terms of the field–dependent one–electron density function $\rho(\boldsymbol{r},\boldsymbol{F})$:[7–9]

$$\mu_i(\boldsymbol{0}) = -\int r_i \rho(\boldsymbol{r},\boldsymbol{0}) d\boldsymbol{r}$$

$$\alpha_{ij} = -\int r_i \rho^{(j)}(\boldsymbol{r}) d\boldsymbol{r} \quad ; \quad \rho^{(j)}(\boldsymbol{r}) = \left.\frac{\partial \rho(\boldsymbol{r},\boldsymbol{F})}{\partial F_j}\right|_{\boldsymbol{F}=\boldsymbol{0}} \quad (5)$$

$$\beta_{ijk} = -\int r_i \rho^{(jk)}(\boldsymbol{r}) d\boldsymbol{r} \quad ; \quad \rho^{(jk)}(\boldsymbol{r}) = \left.\frac{\partial^2 \rho(\boldsymbol{r},\boldsymbol{F})}{\partial F_j \partial F_k}\right|_{\boldsymbol{F}=\boldsymbol{0}} \quad (6)$$

$$\gamma_{ijkl} = -\int r_i \rho^{(jkl)}(\boldsymbol{r}) d\boldsymbol{r} \quad ; \quad \rho^{(jkl)}(\boldsymbol{r}) = \left.\frac{\partial^3 \rho(\boldsymbol{r},\boldsymbol{F})}{\partial F_j \partial F_k \partial F_l}\right|_{\boldsymbol{F}=\boldsymbol{0}} \quad (7)$$

Although the aforementioned density derivatives $\rho^{(j)}(\boldsymbol{r})$, $\rho^{(jk)}(\boldsymbol{r})$ and $\rho^{(jkl)}(\boldsymbol{r})$ have been named after their NLOPs predecessors: $\alpha$–, $\beta$–, $\gamma$–densities, respectively,[8,9] they are not rigorous property densities since their integration does not give the corresponding NLOP.

In a given basis (for instance atomic orbital (AO) basis), the induced dipole moment $\mu_i(\boldsymbol{F})$ can also be defined in terms of the

field–dependent one–particle reduced density matrix (1-RDM), D($\bm{F}$), and one–electron transition dipole moment matrix $\mathbf{h}^{(i)}$ (with elements defined as $h^{(i)}_{\mu\nu} = \langle \phi_\mu | r_i | \phi_\nu \rangle$):

$$\mu_i(\bm{F}) = -\sum_{\mu\nu} D_{\mu\nu}(\bm{F}) h^{(i)}_{\nu\mu} + \sum_A Z_A R_{iA} \quad (8)$$

Further differentiation with respect to the external electric field leads to similar expressions for linear and NLOPs (note that the basis is unchanged for the perturbed and unperturbed systems, i.e., $\partial^n h^{(i)}_{\nu\mu}/\partial F^n_j = 0$):[44,45]

$$\alpha_{ij} = -\sum_{\mu\nu} D^{(j)}_{\mu\nu} h^{(i)}_{\nu\mu} \quad (9)$$

$$\beta_{ijk} = -\sum_{\mu\nu} D^{(jk)}_{\mu\nu} h^{(i)}_{\nu\mu} \quad (10)$$

$$\gamma_{ijkl} = -\sum_{\mu\nu} D^{(jkl)}_{\mu\nu} h^{(i)}_{\nu\mu} \quad (11)$$

In the latter equations, $D^{(j)}_{\mu\nu}$, $D^{(jk)}_{\mu\nu}$ and $D^{(jkl)}_{\mu\nu}$ are the derivatives of 1-RDMs with respect to the external field (expressed in the same basis), which employ the equivalent notation to the one from Equations 5-7.

## 3 Partition of NLOPs into Orbital Contributions

In this section we describe the partition of NLOPs into orbital contributions (PNOC), which is based on the 1-RDM representation of the NLOPs. Only the PNOC expressions for the second hyperpolarizability component $\gamma_{ijkl}$ are given. An equivalent derivation and partition applied to $\alpha$ and $\beta$ is available in the ESI.†

Firstly, the electric field derivative of 1-RDM, $\mathbf{D}^{(jkl)}$, is computed. In the current PNOC implementation, $\mathbf{D}^{(jkl)}$ are calculated with the finite field procedure using the AOs basis, because it is invariant to the applied electric perturbation.

Secondly, the $\mathbf{h}^{(i)}$ and $\mathbf{D}^{(jkl)}$ matrices in the atomic orbital (AO) basis are projected to matrices $\mathbf{M}^{(i)}$ and $\mathbf{\Delta}^{(jkl)}$ that are expressed in the basis of the natural orbitals (NOs) of the unperturbed (i.e., field–free) molecule*, giving rise to an equivalent representation of $\gamma_{ijkl}$ in the new basis,

$$\gamma_{ijkl} = -\sum_{\mu\nu} D^{(jkl)}_{\mu\nu} h^{(i)}_{\nu\mu} = -\text{Tr}(\mathbf{D}^{(jkl)} \mathbf{h}^{(i)})$$

$$= -\text{Tr}(\mathbf{C}^{-1} \mathbf{D}^{(jkl)} (\mathbf{C}^{-1})^\dagger \mathbf{C}^\dagger \mathbf{h}^{(i)} \mathbf{C})$$

$$= -\text{Tr}(\mathbf{\Delta}^{(jkl)} \mathbf{M}^{(i)}) = -\sum_{pq} \Delta^{(jkl)}_{pq} M^{(i)}_{qp}, \quad (12)$$

with the new matrices $\mathbf{M}^{(i)} = \mathbf{C}^\dagger \mathbf{h}^{(i)} \mathbf{C}$ and $\mathbf{\Delta}^{(jkl)} = \mathbf{C}^{-1} \mathbf{D}^{(jkl)} (\mathbf{C}^{-1})^\dagger$, obtained through the transformation between two bases

$$\bm{\chi}^{NO} = \bm{\phi}^{AO} \mathbf{C}, \quad (13)$$

---

*In such representation, $M^{(i)}_{pq} = \langle \chi_p | r_i | \chi_q \rangle$ is the $i$–th component of the transition dipole moment vector between the $p$–th and $q$–th natural orbitals.

where $\bm{\chi}^{NO}$ and $\bm{\phi}^{AO}$ are row vectors, and the LCAO coefficients are organized in columns of matrix $\mathbf{C}$.

Finally, the selected property is partitioned into $p$ NOs components, $\gamma_{ijkl,p}$. In PNOC, each $\gamma_{ijkl,p}$ component is obtained under the assumption that the pair contribution of two NOs $p$ and $q$ is equally distributed between them:

$$\gamma_{ijkl} = -\sum_{pq} \Delta^{(jkl)}_{pq} M^{(i)}_{qp} = \sum_p \gamma_{ijkl,p} \quad (14)$$

$$\gamma_{ijkl,p} = -\Delta^{(jkl)}_{pp} M^{(i)}_{pp} - \frac{1}{2} \sum_{q \neq p} \left( \Delta^{(jkl)}_{pq} M^{(i)}_{qp} + \Delta^{(jkl)}_{qp} M^{(i)}_{pq} \right)$$

$$= -\Delta^{(jkl)}_{pp} M^{(i)}_{pp} - \sum_{q \neq p} \Delta^{(jkl)}_{pq} M^{(i)}_{pq} \quad (15)$$

Although PNOC can use any arbitrary basis of orbitals, we employ NOs of the unpertubed system for the sake of convenience, as this orbital basis is often employed in the description of the 1-RDM of many-body wavefunctions. In the case of spin unrestricted computations, such as UHF (or broken spin–symmetry KS–DFT) we employ the unrestricted NOs defined by Pulay and Hamilton.[46] Indeed, the PNOC can be applied to any quantum chemistry method provided a 1-RDM is available and, unlike other partition methods, it yields an exact decomposition of the NLOP. The main advantages of PNOC are the simplicity and consistency of the partition for the high-order NLOPs.

One of the drawbacks of most of the NLOPs decomposition schemes found in the literature is the so–called *origin–dependency*, which originates from the explicit dependency of dipole moment operator on the position vector.[8,14,15,17,18,25,47] While for the neutral system (total charge $Q$=0) with fixed orientation the total values of the properties (such as $\alpha_{zz}$, $\beta_{zzz}$ and $\gamma_{zzzz}$) do not depend on the relative position with respect to the center of the Cartesian system, the partitioned values in terms of 3D fragments or functions that depend on spatial coordinates usually do. Nevertheless, it is interesting to note that for systems that have the $\sigma_h$ symmetry plane along the target direction or systems that are centrosymmetric, the PNOC decomposition is free of the origin–dependency problem (see Section S.II in ESI† for further details). As previously done by many authors, we adopt here the convention to center the origin at the center of mass of the molecule.[17]

## 4 Computational details and studied systems

In this work we analyze molecular (hyper)polarizabilities of the four different chemical models presented in Figure 1. In all calculations, the center of mass of the system was placed at the origin of the Cartesian coordinate system, and the system was rotated so that the main component of the diagonalized inertia tensor coincides with the $z$–axis. In this arrangement, only the longitudinal components of $\alpha_{zz}$ and $\gamma_{zzzz}$ were studied. Unless stated otherwise, the highest possible symmetry constraints were applied.

The electronic structure calculations were performed with the Gaussian09[48] computational package and the PNOC scheme was implemented within an in–house code.[49] The Chemcraft program was used to generate all the orbital pictures.[50] Our SOS computations include the first 20 (16) FCI (TDHF) $\Sigma_u^+$ excited states ($D_{2h}$ symmetry constraints were applied to the wavefunction). FCI computations were carried out with Molcas8,[51,52] whereas TDHF were done with Gaussian09. In all cases, excepting the $(H_2)_3$ chain (for which we have considered the geometry shown in Figure 1), we have studied the ground state geometry of the molecule. Benzene and $p$–benzyne were optimized using (U)B3LYP/aug–cc–pVDZ (the same basis set was employed for NLOP calculations), whereas for all–*trans*–hexatriene we employed the CAM–B3LYP/cc–pVDZ method. For the $(H_2)_3$ computations we employed the cc–pVDZ basis set of Dunning *et al.*.[53] This basis was chosen because its size is small enough to permit the FCI calculations of $(H_2)_3$ (6 electrons in 30 active orbitals), which is the only electronic structure method for which the SOS expressions yield the exact optical response.

In Section 5.4, we test performance of different basis sets in the calculation of NLOPs of all-trans-hexatriene. Starting from cc–pVDZ/cc–pVDZ for carbon/hydrogen (abbreviated as VDZ/VDZ) we add diffuse functions of selected angular momentum (*s*, *p* or *d* type, with the original primitive exponents) to the basis set.[54] The addition of diffuse functions is represented with +. Following this nomenclature, the basis sets that have been tested (total number of basis functions is given in parentheses) are: VDZ/VDZ (124), VDZ+s/VDZ (130), VDZ+p/VDZ (142), VDZ+d/VDZ (154), VDZ+sp/VDZ (148), VDZ+p/VDZ+p (166), AVDZ/VDZ (178), VDZ/AVDZ (156) and AVDZ/AVDZ (210). Note that with such notation, the AVDZ basis set is equivalent to VDZ+spd for carbon and to VDZ+sp for hydrogen.

To achieve the best stability in the 1-RDMs numerical differentiation (see Section S.III in ESI†), tight convergence criteria were used in the electronic structure computations (given in atomic units): $10^{-11}$ in HF, $10^{-11}$ in CASSCF total energies, $10^{-10}$ and $10^{-8}$ for the energy and the norm of the amplitudes vector in CCSD. Unrestricted coupled Hartree-Fock (UCHF) calculations were obtained from the data computed with Gaussian09 using the expressions given in section S.IV of the ESI†.

## 5 Results and discussion

### 5.1 Comparison between SOS and PNOC scheme – linear $(H_2)_3$ chain

In this section, we demonstrate the potential of the PNOC decomposition by comparing the NOs contributions to the ones provided by SOS (for a description of the SOS see Section S.IV of ESI†). To this end, we analyze $\alpha_{zz}$ of a model system of three colinear hydrogen molecules, $(H_2)_3$ (see Figure 1). As expected, the SOS and the PNOC FCI values of the NLOPs are almost equal. A small negligible difference of 0.12 a.u. for $\alpha_{zz}$ originates from the incompleteness of the set of FCI excited states used in the SOS formulation (see Table 1). For this system, $\alpha_{zz}^{SOS}$ using TDHF excited states is also in excellent agreement with the coupled HF value of $\alpha_{zz}$. Such good agreement between TDHF SOS and coupled NLOPs is in general not expected for larger molecular systems. On the other hand, UCHF $\alpha_{zz}^{SOS}$ is underestimated by 34.5% with respect to the coupled HF value.

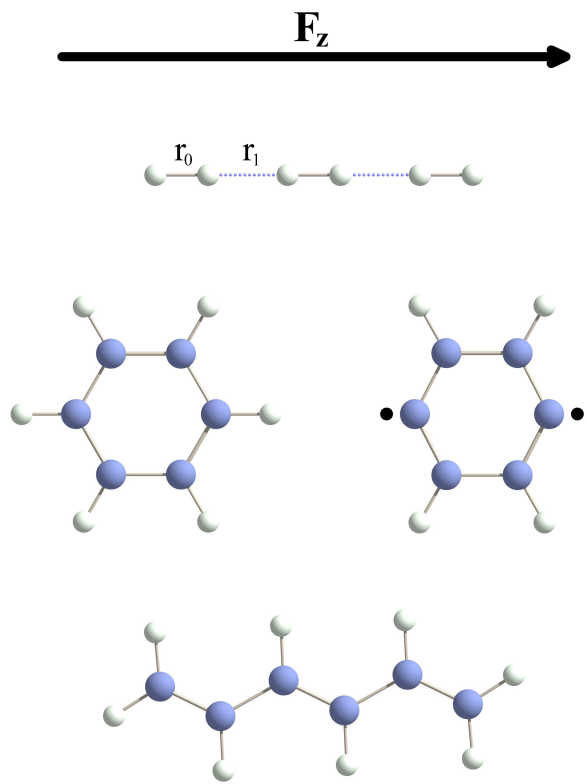

**Fig. 1** Studied model chemical systems and the direction of the applied external electric field $F_z$. *Top:* $(H_2)_3$ chain with intramolecular $r_0$=2 bohr and intermolecular $r_1$=3 bohr; *Center:* Benzene (left) and *p*–benzyne (right); *Bottom:* All–*trans*–hexatriene.

Results of the SOS orbital partitioning and the PNOC scheme are compared in Table 1. In general, a semiquantitative agreement between SOS and PNOC is observed. The orbital contributions to the NLOP provided by the SOS and PNOC schemes are both in agreement with chemical intuition: the frontier orbitals, $2\sigma_g^+$ (HOMO) and $2\sigma_u^+$ (LUMO) in $(H_2)_3$ are the most important to the linear response (see Figure S.1 in the ESI† for a graphical representation of all NOs along with their occupancies). The relative contributions of the NOs to SOS and PNOC coupled $\alpha_{zz}$ are very similar (differences smaller than 11 %) – especially for the HF wavefunction, for which exactly the same set of MOs of the ground state is used in both decompositions (differences smaller than 6 %). The differences between the SOS and PNOC contributions to FCI $\alpha_{zz}$ originate from the approximate SOS orbital partitioning, which assumes that the NOs of the excited states are the same as those of the ground state. Conversely, in PNOC, this assumption is not required since the shape and occupancies of the perturbed NOs are projected to the NOs of the unperturbed system. It is interesting to notice that whereas the PNOC decompositions of static first and second hyperpolarizabilities are as simple as the PNOC splitting of the linear polarizability, the complexity of the SOS analysis rapidly increases with the order of property.

## 5.2 Orbital contributions to NLOPs in benzene and p–benzyne

In this section, we evaluate PNOC as diagnostic tool for the performance of quantum chemical methods in the NLOPs computations. This is done for two simple chemical systems: benzene ($C_6H_6$), a typical closed–shell aromatic system, and p–benzyne ($C_6H_4$), a diradical molecule in its singlet open–shell state with two unpaired electrons in *para* position (see Figure 1).[55]

We focus on (*i*) the changes in the chemical nature of the response arising from the formation of the diradical in *para* position (analysis based on (U)CCSD results, data compiled in Table 2), and (*ii*) the differences in the description of (non)linear response by (U)HF, (U)MP2, CASSCF, and (U)CCSD methods (see Tables 2, 3, and 4). We analyze the contributions of the following sets of orbitals of $C_6H_6$ [$D_{6h}$] ($C_6H_4$ [$D_{2h}$]): $1s_c$, carbon core orbitals; $\sigma$(occ), $\sigma$–type orbitals lying below the Fermi level (orbitals $6e_{1u}$ ($5b_{1u}$) and $6e_{2g}$ ($6a_g$) belong to this group); $\sigma$(vir), $\sigma$–type orbitals lying above the Fermi level, $\pi$(occ), $\pi$–type lying below the Fermi level (orbitals $1a_{2u}$ ($1b_{3u}$), $1e_{1g}$ ($1b_{1g}$) and $2e_{1g}$ ($1b_{2g}$)); $\pi$(vir), $\pi$–type lying above the Fermi level (orbitals $1e_{2u}$ ($2b_{3u}$), $2e_{2u}$ ($1a_u$) and $1b_{2g}$ ($2b_{2g}$)). In benzene, valence $\pi$–type orbitals (three occupied: $1a_{2u}$, $1e_{1g}$, $2e_{1g}$, and three virtual: $1e_{2u}$, $2e_{2u}$, $1b_{2g}$) drive most of the properties, whereas in p–benzyne two additional orbitals of $\sigma$–type, $5b_{1u}$ and $6a_g$ (corresponding to the two unpaired electrons) are also relevant (see Figure S.2 in ESI† for their graphical representation). Therefore, we performed CASSCF(6,6) calculations for benzene and CASSCF(8,8) for p–benzyne, with active spaces defined by these NOs.

In our analysis, we choose the (U)CCSD orbital contributions as the reference values to assess the PNOC analysis at lower levels of theory. The total values of NLOPs computed using (U)CCSD(T) are also given for comparison.

According to PNOC, the difference between the overall contributions of $\pi$– and $\sigma$–type NOs to the linear response of benzene is smaller than 3% (see Table 2). As expected, the core orbitals are not involved in the static response (core electrons are highly confined near the nuclei). The largest contributions to CCSD $\alpha_{zz}$ correspond to four frontier $\pi$ NOs: $1e_{1g}$, $2e_{1g}$, $1e_{2u}$ and $2e_{2u}$.

In the third-order response, the main role is shifted towards the $\pi$–type orbitals, for which the overall contribution to $\gamma_{zzzz}$ is about 60%. The most important contributions to $\alpha_{zz}$ come from cross-orbitals terms mixing occupied and virtual orbitals, whereas the dominant contributions to $\gamma_{zzzz}$ are due to terms involving only virtual orbitals. Indeed, the largest individual contributions belong to the frontier $2e_{1g}$ and $1e_{2u}$ orbitals, whereas the relative contribution of $1e_{1g}$ and $2e_{2u}$ is not as high as in the case of $\alpha_{zz}$. All occupied $\sigma$ orbitals along with $1e_{1g}$ and $2e_{2u}$ $\pi$ orbitals have a small negative contribution to the property. Indeed the largest contributions to $\gamma_{zzzz}$ come from the overall virtual $\sigma$ and $\pi$ orbitals. However, closer inspection reveals that individual higher virtual orbitals do not have large contributions, but there is a large number of small contributions adding to the total.

The formation of the singlet p–benzyne diradical changes the character of the electronic structure of the system, which has two unpaired electrons (of opposite spin) occupying $5b_{1u}$ and $6a_g$ NOs with the occupancies 1.18 and 0.81, respectively (see Figure S.1 in ESI†). Such phenomenon is a direct manifestation of *type a* nondynamic correlation,[56] which results from an absolute near–degeneracy of HOMO ($5b_{1u}$) and LUMO ($6a_g$) orbitals. In single–reference HF and post–HF methods (such as MP2 or CCSD), symmetry breaking allows to partially account for this type of electron correlation.[56] Usually this correlation effect has huge impact,[57] especially on higher order NLOPs.[11,58–62]

The formation of $C_6H_4$ diradical from $C_6H_6$ decreases the longitudinal CCSD $\alpha_{zz}$ by approximately 5.3 a.u. However, the PNOC analysis reveals that in fact the contribution of the occupied $\sigma$–type NOs increases by 6.0 a.u., and the lowering of the linear response is observed due to the virtual $\sigma$ and occupied and virtual $\pi$–type orbitals. Within the latter group, the largest decrease is reported for the frontier $\pi$–type NOs, *i.e.*, $2e_{1g}$ ($1b_{2g}$) and $1e_{2u}$ ($2b_{3u}$), respectively.

In contrast to the small decrease of $\alpha_{zz}$ upon formation of the diradical, the total value of CCSD $\gamma_{zzzz}$ dramatically increases by 17148 a.u (140%). The PNOC decomposition shows that the nature of the response in p–benzyne ($\gamma_{zzzz}$) is completely dominated by the cross-orbital terms corresponding to two radical orbitals, $5b_{1u}$ and $6a_g$ (the contribution of both NOs is 46% of the total $\gamma_{zzzz}$). Unlike $C_6H_6$, $C_6H_4$ $\gamma_{zzzz}$ is almost fully dominated by the response of the $\sigma$–type NOs, and, although the larger overall part corresponds to the $\sigma$(vir), for $C_6H_4$ the $\sigma$(occ) also plays a very important role. Conversely, $\pi$–type orbitals display a negative small contribution, coming mostly from the valence $\pi$ NOs (see Table 2).

Next, we discuss the capability of PNOC to decompose the effect of the electron correlation in the description of static NLOPs. On the examples of benzene and p–benzyne, we analyze the effect of the electronic correlation through the differences in the NO contributions to the properties obtained with the HF method and selected post–HF methods: MP2 and CCSD (which introduce dynamic correlation) and CASSCF (which introduces nondynamic correlation). HF, MP2 and CASSCF are compiled in Tables 3 and 4, whereas CCSD values, which serve as the reference, are taken from Table 2.

In benzene, the total $\alpha_{zz}$ obtained from UHF is almost equal to the CCSD one. MP2 overestimates it by 2.0 a.u., whereas CASSCF underestimates it by 5.1 a.u. This suggests that UHF describes the linear response better than CASSCF, which by the variational principle yields energies closer to the exact value. PNOC reveals that CASSCF underestimates $\alpha_{zz}$ due to a systematically smaller contribution of the $\pi$–type orbitals to $\alpha_{zz}$. On the other hand, MP2, which overestimates total $\alpha_{zz}$, actually describes the contributions for each of the $\sigma$ and $\pi$ valence orbitals much better than HF or CASSCF. The small error of HF total $\alpha_{zz}$ is actually due to a fortunate compensation between an underestimation of the contribution of the $\sigma$ orbitals and an overestimation of the contribution of the $\pi$ orbitals.

In comparison to $C_6H_6$, the total value of $\alpha_{zz}$ for $C_6H_4$ is smaller by 15.6 a.u. in HF, 1.4 a.u. in MP2, and 7.4 a.u. in CASSCF (with respect to the CCSD value of 5.3 a.u.). Although the magnitude of the MP2 and CASSCF relative error of $\alpha_{zz}$ with respect to UCCSD value is similar, changes in the nature of the linear response, such as the increase of the contribution of 5b$_{1u}$ (and decrease of 6a$_g$) and overall decrease in contributions of the $\pi$–type NOs (especially frontier orbitals) are described better by CASSCF than UHF and MP2. UMP2 significantly overestimates the relative contribution of 1b$_{1g}$ and 1a$_u$ NOs to the total $\alpha_{zz}$. The inaccuracy given by UHF and UMP2 partition of the linear response is in agreement with the important role of nondynamic correlation in $p$–benzyne.

**Table 3** NOs contributions to longitudinal $\alpha_{zz}$ (in a.u.) of benzene (C$_6$H$_6$) and $p$–benzyne (C$_6$H$_4$) obtained at different levels of theory. In the electronic structure computations, restricted and unrestricted HF/MP2 variants were used for C$_6$H$_6$ and C$_6$H$_4$, whereas in the CASSCF/aug–cc–pVDZ computations active spaces of (6,6) and (8,8) were selected, respectively.

|  | (U)HF | | (U)MP2 | | CASSCF | |
|---|---|---|---|---|---|---|
| Orbitals | C$_6$H$_6$ | C$_6$H$_4$ | C$_6$H$_6$ | C$_6$H$_4$ | C$_6$H$_6$ | C$_6$H$_4$ |
| 6e$_{1u}$ (5b$_{1u}$) | 4.43 | 4.67 | 4.31 | 6.56 | 4.58 | 6.70 |
| 6e$_{2g}$ (6a$_g$) | 4.67 | 2.70 | 4.21 | 3.38 | 4.93 | 3.76 |
| 1a$_{2u}$ (1b$_{3u}$) | 0.38 | 1.06 | 0.74 | 0.29 | 0.81 | 0.84 |
| 1e$_{1g}$ (1b$_{1g}$) | 9.50 | 6.93 | 8.60 | 12.96 | 7.44 | 7.30 |
| 2e$_{1g}$ (1b$_{2g}$) | 12.07 | 7.05 | 11.24 | 9.13 | 10.01 | 6.91 |
| 1e$_{2u}$ (2b$_{3u}$) | 8.42 | 4.17 | 7.59 | 6.02 | 6.72 | 4.18 |
| 2e$_{2u}$ (1a$_u$) | 8.42 | 6.60 | 7.29 | 11.58 | 6.63 | 6.51 |
| 1b$_{2g}$ (2b$_{2g}$) | 0.00 | 0.12 | 0.38 | 0.01 | 0.25 | 0.19 |
| 1s$_c$ | 0.01 | 0.01 | 0.01 | 0.01 | 0.07 | 0.01 |
| $\sigma$(occ) | 18.08 | 19.97 | 18.41 | 21.12 | 19.00 | 22.65 |
| $\sigma$(vir) | 18.09 | 14.55 | 20.84 | 14.72 | 19.01 | 14.59 |
| $\sigma$(all) | 36.18 | 34.52 | 39.25 | 35.84 | 38.01 | 37.24 |
| $\pi$(occ) | 21.96 | 15.03 | 20.58 | 22.39 | 18.26 | 15.04 |
| $\pi$(vir) | 21.96 | 14.90 | 22.04 | 22.22 | 18.43 | 14.98 |
| $\pi$(all) | 43.92 | 29.93 | 42.62 | 44.60 | 36.68 | 30.02 |
| Total $\alpha_{zz}$ | 80.10 | 64.47 | 81.89 | 80.45 | 74.70 | 67.27 |

Differences in the description of the response are especially large for the second hyperpolarizability. In benzene, the $\gamma_{zzzz}$ total values obtained with HF and CASSCF are underestimated (with respect to CCSD ones). This drawback is found neither for MP2 or CCSD, which include dynamic correlation and provide a very similar value of $\gamma_{zzzz}$. The PNOC also supports this conclusion: MP2 provides a well–balance description of the nonlinear response coming from $\sigma$, as well as $\pi$ NOs (both occupied and virtual), Conversely, CASSCF strongly underestimates the overall contribution of the $\sigma$–type NOs and yields very inaccurate individual contributions of the valence $\pi$ 1e$_{1g}$ and 2e$_{2u}$ NOs, revealing that the nonlinear response given by CAS(6,6) is still incorrect.

Upon formation of the diradical $p$–benzyne, dramatic discrepancies in the nature of $\gamma_{zzzz}$ are observed among these four methods. Interestingly, PNOC detailed analysis does not follow the trends expected from the total values of $\gamma_{zzzz}$, i.e., in comparison to UCCSD, UMP2 provides the best total value for the nonlinear response, while UHF and CASSCF(8,8) perform unsatisfactory. However, PNOC analysis unveils that the small error of the MP2 total $\gamma_{zzzz}$ value of $p$–benzyne is accidental and comes from a fortunate cancellation of large errors. The largest discrepancy is observed for the response given by the 5b$_{1u}$ and 6a$_g$ NOs (the only ones occupied by radical electrons), which incorrectly display an insignificant role in the nonlinear response. Similar errors are also observed for UHF decomposed NLOPs. Furthermore, according to the reference UCCSD results, relative contributions of $\pi$(all) should be -3.7% which UHF and UMP2 greatly overestimate (49.9% by UHF and 40.0% by UMP2). In contrast, the role of 5b$_{1u}$ and 6a$_g$ NOs is correctly described by the CASSCF(8,8) calculation. These two orbitals contribute approximately to 50% of the total CASSCF $\gamma_{zzzz}$ (against 46% in UCCSD). Moreover, at the CASSCF level of theory, the response of $\pi$ NOs has dramatically decreased in the diradical (as compared to benzene) down to 8.3%, which is in much better agreement with UCCSD predictions. PNOC analysis shows that the error of CASSCF $\gamma_{zzzz}$ is again due to a strong (but systematic) underestimation of the overall contribution of the virtual $\sigma$–type NOs.

**Table 4** NOs contributions to longitudinal $\gamma_{zzzz}$ (in a.u.) of benzene (C$_6$H$_6$) and $p$–benzyne (C$_6$H$_4$) obtained at different levels of theory. In the electronic structure calculations, restricted and unrestricted HF/MP2 variants were used for C$_6$H$_6$ and C$_6$H$_4$, whereas in the CASSCF/aug–cc–pVDZ computations active spaces of (6,6) and (8,8) were selected, respectively.

|  | (U)HF | | (U)MP2 | | CASSCF | |
|---|---|---|---|---|---|---|
| Orbitals | C$_6$H$_6$ | C$_6$H$_4$ | C$_6$H$_6$ | C$_6$H$_4$ | C$_6$H$_6$ | C$_6$H$_4$ |
| 6e$_{1u}$ (5b$_{1u}$) | -377 | 733 | -338 | 2796 | -377 | 6544 |
| 6e$_{2g}$ (6a$_g$) | -157 | -288 | 75 | 364 | -229 | 3683 |
| 1a$_{2u}$ (1b$_{3u}$) | 311 | 45 | 119 | -558 | 269 | 103 |
| 1e$_{1g}$ (1b$_{1g}$) | -1300 | 1983 | -493 | 2109 | 486 | 43 |
| 2e$_{1g}$ (1b$_{2g}$) | 1555 | 931 | 686 | 1523 | 1706 | -41 |
| 1e$_{2u}$ (2b$_{3u}$) | 3475 | 843 | 296 | 1694 | 1363 | -65 |
| 2e$_{2u}$ (1a$_u$) | -396 | 1688 | -375 | 1866 | 411 | 22 |
| 1b$_{2g}$ (2b$_{2g}$) | 411 | 154 | 423 | 211 | 391 | 137 |
| 1s$_c$ | 0 | -1 | 0 | -1 | 0 | -2 |
| $\sigma$(occ) | -1018 | 35 | -987 | 2457 | -1246 | 9113 |
| $\sigma$(vir) | 4305 | 7559 | 6739 | 13542 | 4327 | 10140 |
| $\sigma$(all) | 3287 | 7594 | 5752 | 15999 | 3080 | 19253 |
| $\pi$(occ) | 567 | 2959 | 311 | 3074 | 2461 | 105 |
| $\pi$(vir) | 5352 | 4610 | 6089 | 7612 | 5488 | 1634 |
| $\pi$(all) | 5919 | 7568 | 6400 | 10686 | 7948 | 1739 |
| Total $\gamma_{zzzz}$ | 9206 | 15162 | 12152 | 26683 | 11028 | 20990 |

### 5.3 Cartesian representation of NLOPs using PNOC

Another useful feature of PNOC is its capability of representing the NLOPs in the Cartesian space without any additional computational cost. Within the PNOC scheme, a local representation of the NLOPs can be obtained through multiplication of each NO contribution to the property, i.e.. $\alpha_{zz,p}$ or $\gamma_{zzzz,p}$ by its NO local amplitude, $|\chi_p(\bm{r})|^2$:

$$\alpha_{zz}(\bm{r}) = \sum_p \alpha_{zz,p} |\chi_p(\bm{r})|^2 \qquad (16)$$

$$\gamma_{zzzz}(\bm{r}) = \sum_p \gamma_{zzzz,p} |\chi_p(\bm{r})|^2 \qquad (17)$$

This approach tacitly assumes a uniform contribution of the orbital to the property, as some of us have done in other contexts.[55] One of the most important properties of the latter definitions of $\alpha_{zz}(r)$ and $\gamma_{zzzz}(r)$, is that the total values of $\alpha_{zz}$ and $\gamma_{zzzz}$ are retrieved upon integration over the whole Cartesian space (likewise, the integration of the local $p$–th NO contribution yields the total NO contribution due to orbital $p$). This conclusion does not hold for density derivatives $\rho^{(1)}(r)$ and $\rho^{(3)}(r)$, which always integrate to zero.

In Figure 2, comparison of the information provided by $\alpha_{zz}(r)$ and $\gamma_{zzzz}(r)$, and density derivatives $\rho^{(1)}(r)$ and $\rho^{(3)}(r)$ is shown for benzene and $p$–benzyne. Overall, the pictures provided by PNOC and density derivatives are in good agreement. However, PNOC representations are simpler and easier to analyze. For example, in $C_6H_6$, all carbons contribute equally to the polarizability and much more than hydrogens. Among those, the hydrogens in *para* position have slightly larger contribution to the longitudinal polarizability. $\gamma_{zzzz}(r)$ is mostly localized in the vicinity of the C1 and C4 atoms, and is showing a predominant $\pi$ character. In $C_6H_4$, carbons with unpaired electrons have larger contribution to both the linear and nonlinear response, although the localization is far larger in $\gamma_{zzzz}$. Additionally, for $\gamma_{zzzz}(r)$, one can see the small negative contributions of $\pi$ type of orbitals at carbon atoms without unpaired electrons, in agreement with the results presented in Section 5.2.

Interestingly, a good qualitative correspondence is found between local NLOPs and the profiles of local nondynamic ($I_{nd}(r)$) and dynamic ($I_d(r)$) electron correlation of the unperturbed system.[55] This can be easily seen in $p$–benzyne diradical, for which the regions with large $I_{nd}(r)$ approximately match $\gamma_{zzzz}(r)$. On the other hand, in benzene, $\gamma_{zzzz}(r)$ follows the profile of $I_d(r)$, but with larger contributions from the carbons aligned in the direction of the applied field.

### 5.4 Basis set dependency

PNOC proves useful also in the benchmark studies of the basis set effect in the NLOPs computations. In this section, we study the influence of diffuse functions in the character of the response using the PNOC analysis. We analyze a small $\pi$–conjugated system, all–*trans*–hexatriene ($C_6H_8$), calculated with the CCSD method. Starting from the cc–pVDZ basis set, we separately add diffuse functions of selected angular momentum to the basis functions of carbon (C) or/and hydrogen (H), (see Section 4 for the description of these basis sets). We analyze the overall contributions of NOs belonging to two defined subspaces – full valence (FV) and higher virtual (HV) orbitals. The FV subspace consists of first 19 occupied (including core $1s_C$ orbitals) and first 16 virtual $\sigma$–type NOs, and 3 occupied and first 3 virtual $\pi$–type NOs. All remaining virtual $\sigma$ and $\pi$ NOs are classified as the HV orbitals (and their number depends on the size of the basis set).

Results of calculated and decomposed NLOPs are compiled in Tables 5 and 6. PNOC reveals that the addition of diffuse functions to the basis sets increase the values of polarizability and second hyperpolarizability in a different manner. Changing from VDZ/VDZ to AVDZ/AVDZ, $\alpha_{zz}$ rises only by 10%, 52% of this in-crement coming from the contributions of the valence NOs (and the rest from the HV NOs). On the other hand, large changes of $\gamma_{zzzz}$ in hexatriene are mainly due to the larger contributions of the HV orbitals (which account for 72% of the property value). As expected, the second–order hyperpolarizability is more basis-set dependent than the polarizability.

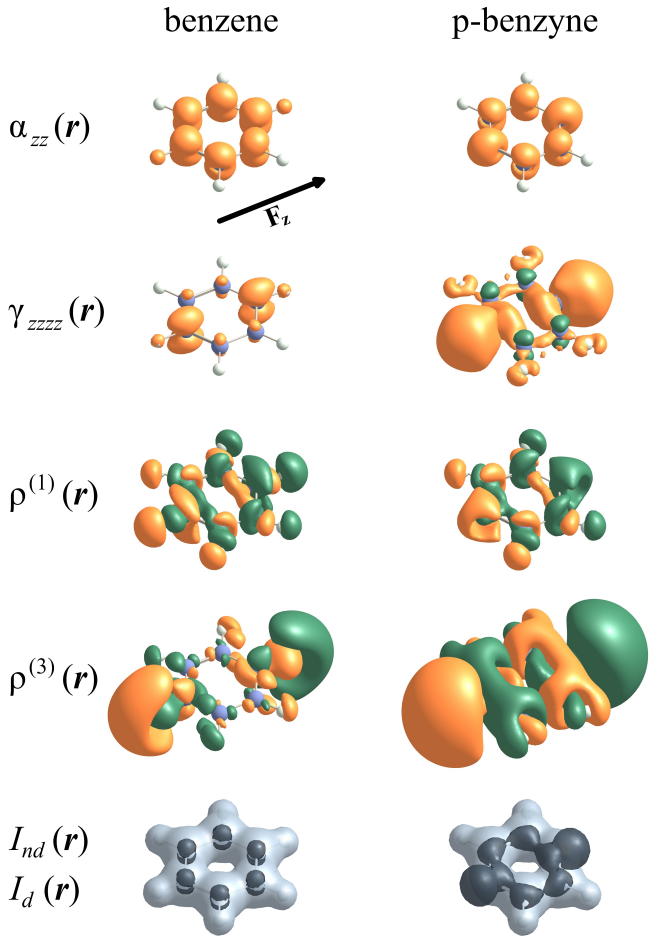

**Fig. 2** Comparison of functions representing the NLOPs in the Cartesian space for benzene (left) and $p$–benzyne (right): $\alpha_{zz}(r)$ and $\gamma_{zzzz}(r)$ obtained from PNOC contributions, and density derivatives $\rho^{(1)}(r)$ and $\rho^{(3)}(r)$. Orange and green colors represent positive and negative values of the selected isosurfaces. The last row presents local nondynamic and dynamic correlation indices $I_{nd}(r)$ (light gray) and $I_d(r)$ (dark gray),[55] respectively. All the representations were obtained at the (U)CCSD/aug–cc–pVDZ level of theory. We have used the following isosurface values (in atomic units): $\alpha_{zz}(r) = 0.40$, $\gamma_{zzzz}(r) = 40.0$, $\rho^{(1)}(r) = 0.085$, $\rho^{(3)}(r) = 5.00$, and $I_{nd}(r) = I_d(r) = 0.006$.

The PNOC analysis can be used to make a judicious selection of the smallest basis set providing an accurate description of NLOPs. In the following example we aim at reproducing the PNOC analysis and the total values of the response calculated with AVDZ/AVDZ basis, with the smallest number of diffuse functions. The data compiled is in Tables 5 and 6. The addition of diffuse functions to carbon atoms is much more important than to the hydrogen centers for the correct description of both $\alpha_{zz}$ and $\gamma_{zzzz}$. Using AVDZ/VDZ basis, a very good agreement with AVDZ/AVDZ is found not only in the total values of $\alpha_{zz}$ and $\gamma_{zzzz}$, but also in the particular orbital components, namely $\sigma$(FV), $\sigma$(HV), $\pi$(FV) and $\pi$(HV). On the other hand, VDZ/AVDZ suffers mostly from the incorrect description of the response of the $\pi$–type NOs, which is especially important for the accurate description of $\gamma_{zzzz}$.

Among the set of diffuse functions of C, $p$–type diffuse functions seem to be the most important for the description of both $\alpha_{zz}$ and $\gamma_{zzzz}$. They substantially enhance the response of the $\pi$–type NOs, especially the part coming from the HV NOs. On the other hand, $s$–type diffuse functions marginally affect the optical response in hexatriene (compare columns 2 and 3 or 4 and 6 in Tables 5 and 6). Interestingly, $d$–type diffuse functions enhance the total value of $\alpha_{zz}$ by almost the same value as the $p$–type functions. However, PNOC unveils that the $\pi$(all) contribution to $\alpha_{zz}$ is not so large, and this drawback is even magnified in $\gamma_{zzzz}$.

Since the diffuse functions of $p$ angular momentum in carbon seem to be crucial, we analyze the effect of adding diffuse functions to H centers too. The VDZ+p/VDZ+p basis set has substantially better performance than VDZ+p/VDZ and yields results comparable to AVDZ/VDZ. In the description of $\alpha_{zz}$, VDZ+p/VDZ+p corrects all orbital contributions to the response that were insufficiently described by VDZ+p/VDZ. The VDZ+p/VDZ+p basis correctly describes the particular $\gamma_{zzzz}$ components of both FV and HV NOs. Needles to say, the reduced size of this basis set could become in handy for the description of NLOPs in large molecules.

## 6 Conclusions

In this paper, we have presented an orbital decomposition scheme for optical properties, PNOC. This tool can be used to obtain a local representation of the NLOPs, providing a powerful visualization aid to connect the magnitude of the optical property with some parts of the molecule. This feature can be employed in the quest for nonlinear optical materials.

Perhaps the most interesting property of PNOC is the fact that the optical properties are decomposed in terms of the natural orbitals of the unperturbed system, making it a convenient method to detect flaws in NLOP evaluation of electronic structure methods. In particular, we have seen how the orbital contributions are more sensible to the choice of the electronic structure method than the total value of the optical property. In this way, PNOC unveils compensations of errors between orbital contributions and helps identifying orbitals that have a prominent role in the description of the property and, therefore, should be included in the active space of orbitals in a correlated calculation. For $p$–benzyne, PNOC has been also used to show that CASSCF provides a poor description of the total value $\gamma_{zzzz}$ because it underestimates the virtual $\sigma$ orbital contribution, whereas MP2 calculations yield inaccurate values due to the lack of nondynamic correlation that causes the underestimation of the occupied $\sigma$ orbitals.

Optical properties (especially of high order) are particularly susceptible to the inclusion of diffuse basis sets that significantly increase the computational cost of the calculation. The PNOC can be also used to identify the proper number and type of atomic basis sets required, saving substantial amounts of computer time for the calculation of optical properties in large molecules.

PNOC decomposes *exactly* the optical property regardless the approximation employed in the electronic structure method. Unlike other analysis methods for NLOPs, PNOC does not pose a large computational cost and it can be straightforwardly applied to any response order. For instance, the exact SOS analysis of the second-order hyperpolarizability involves large expressions and it is rarely applied. PNOC only requires the first-order reduced density matrix, which is easily available in most electronic structure methods. PNOC formulae can be easily generalized to evaluate any static or dynamic one-electron property, such as multipole moments and diamagnetic susceptibilities.

## Conflicts of interest

There are no conflicts to declare.

## Acknowledgements

S.P.S. acknowledges the Basque Government for funding through a predoctoral fellowship (PRE 2018 2 0200). This research has been funded by Spanish MINECO/FEDER Projects CTQ2014-52525-P and EUIN2017-88605, the Catalan DIUE 2014SGR931 and IKERBASQUE. The authors acknowledge the computational resources and technical and human support provided by the DIPC and the SGI/IZO-SGIker UPV/EHU.

## Notes and references

**Table 1** Comparison among the orbital contributions to $\alpha_{zz}$ in $(H_2)_3$ obtained from the SOS and the PNOC schemes using various wave functions. For each particular method, the absolute values (in a.u.) of the NOs contributions to $\alpha_{zz}$ are presented (relative contributions given in parentheses). The full table and the orbitals are included in the ESI†(Tables S.I and Figure S1, respectively).

| NO | SOS(UCHF) | SOS(TDHF) | PNOC(HF) | SOS(FCI) | PNOC(FCI) |
|---|---|---|---|---|---|
| $1\sigma_g^+$ | 0.54 ( 1.5%) | 0.39 ( 0.7%) | 0.87 ( 1.6%) | 0.39 ( 0.8%) | 0.85 ( 1.7%) |
| $1\sigma_u^+$ | 2.72 ( 7.4%) | 2.18 ( 3.9%) | 4.40 ( 7.9%) | 1.74 ( 3.5%) | 4.04 ( 8.0%) |
| $2\sigma_g^+$ | 15.01 (41.1%) | 25.32 (45.4%) | 22.64 (40.6%) | 22.66 (45.1%) | 19.91 (39.5%) |
| $2\sigma_u^+$ | 14.96 (41.0%) | 25.47 (45.7%) | 22.60 (40.5%) | 22.63 (45.0%) | 17.48 (34.7%) |
| $3\sigma_g^+$ | 2.66 ( 7.3%) | 2.14 ( 3.8%) | 4.39 ( 7.9%) | 1.41 ( 2.8%) | 3.17 ( 6.3%) |
| $3\sigma_u^+$ | 0.48 ( 1.3%) | 0.20 ( 0.4%) | 0.87 ( 1.6%) | 0.49 ( 1.0%) | 0.71 ( 1.4%) |
| | | | | | |
| Total $\alpha_{zz}$ | 36.54 | 55.78 | 55.80 | 50.27 | 50.39 |

**Table 2** Contributions of selected NOs to longitudinal $\alpha_{zz}$ and $\gamma_{zzzz}$ of benzene ($C_6H_6$) and $p$–benzyne ($C_6H_4$), obtained at the (U)CCSD/aug–cc–pVDZ level of theory. Units are a.u. and the relative contributions are given in parenthesis. These NOs along with their occupancies are represented in the Cartesian space on Figure S.2 in ESI.†

| | $\alpha_{zz}$ | | $\gamma_{zzzz}$ | |
|---|---|---|---|---|
| Orbitals | $C_6H_6$ | $C_6H_4$ | $C_6H_6$ | $C_6H_4$ |
| $6e_{1u}$ ($5b_{1u}$) | 4.36 ( 5.5%) | 7.83 (10.5%) | -373 (-3.0%) | 8536 (29.0%) |
| $6e_{2g}$ ($6a_g$) | 4.57 ( 5.7%) | 4.48 ( 6.0%) | -77 (-0.6%) | 5112 (17.4%) |
| $1a_{2u}$ ($1b_{3u}$) | 0.66 ( 0.8%) | 0.64 ( 0.9%) | 186 ( 1.5%) | -80 (-0.3%) |
| $1e_{1g}$ ($1b_{1g}$) | 8.39 (10.5%) | 8.78 (11.8%) | -411 (-3.4%) | -1387 (-4.7%) |
| $2e_{1g}$ ($1b_{2g}$) | 10.94 (13.7%) | 7.40 ( 9.9%) | 1402 (11.4%) | -540 (-1.8%) |
| $1e_{2u}$ ($2b_{3u}$) | 7.25 ( 9.1%) | 4.58 ( 6.1%) | 978 ( 8.0%) | -489 (-1.7%) |
| $2e_{2u}$ ($1a_u$) | 7.06 ( 8.8%) | 6.88 ( 9.2%) | -343 (-2.8%) | -1120 (-3.8%) |
| $1b_{2g}$ ($2b_{2g}$) | 0.26 ( 0.3%) | 0.20 ( 0.3%) | 451 ( 3.7%) | 103 ( 0.4%) |
| | | | | |
| $1s_c$ | 0.01 ( 0.0%) | 0.01 ( 0.0%) | 0 ( 0.0%) | -2 ( 0.0%) |
| | | | | |
| $\sigma$(occ) | 18.04 (22.6%) | 24.01 (32.2%) | -1153 (-9.4%) | 12602 (42.9%) |
| $\sigma$(vir) | 20.68 (25.9%) | 16.18 (21.7%) | 6018 (49.1%) | 17879 (60.8%) |
| $\sigma$(all) | 38.72 (48.5%) | 40.19 (53.9%) | 4865 (39.7%) | 30481 (103.7%) |
| | | | | |
| $\pi$(occ) | 19.99 (25.0%) | 16.81 (22.5%) | 1177 ( 9.6%) | -2007 (-6.8%) |
| $\pi$(vir) | 21.12 (26.5%) | 17.57 (23.6%) | 6213 (50.7%) | 932 ( 3.2%) |
| $\pi$(all) | 41.11 (51.5%) | 34.39 (46.1%) | 7390 (60.3%) | -1076 (-3.7%) |
| | | | | |
| Total NLOP | | | | |
| CCSD | 79.84 | 74.58 | 12255 | 29403 |
| CCSD(T) | 80.39 | 76.70 | 13180 | 37152 |

**Table 5** Basis set dependency of the NOs contributions to longitudinal $\alpha_{zz}$ of hexatriene. Decomposed values were obtained at the CCSD level of theory. Additionally, CCSD(T) total values are given.

| C: | VDZ | VDZ+s | VDZ+p | VDZ+d | VDZ+sp | VDZ+p | AVDZ | VDZ | AVDZ |
| H: | VDZ | VDZ | VDZ | VDZ | VDZ | VDZ+p | VDZ | AVDZ | AVDZ |
|---|---|---|---|---|---|---|---|---|---|
| $\sigma$(FV) | 36.72 | 36.86 | 36.72 | 37.19 | 36.86 | 37.11 | 37.13 | 37.28 | 37.19 |
| $\sigma$(HV) | 7.79 | 8.06 | 8.23 | 8.84 | 8.42 | 9.01 | 9.08 | 8.99 | 9.22 |
| $\sigma$(all) | 44.51 | 44.92 | 44.95 | 46.03 | 45.28 | 46.12 | 46.21 | 46.28 | 46.41 |
| | | | | | | | | | |
| $\pi$(FV) | 83.50 | 83.62 | 88.18 | 86.84 | 88.27 | 89.51 | 90.57 | 86.63 | 90.84 |
| $\pi$(HV) | 4.92 | 4.93 | 7.46 | 6.99 | 7.50 | 8.45 | 9.01 | 6.80 | 9.20 |
| $\pi$(all) | 88.42 | 88.55 | 95.64 | 93.83 | 95.76 | 97.96 | 99.58 | 93.43 | 100.04 |
| | | | | | | | | | |
| Sum(FV) | 120.22 | 120.48 | 124.90 | 124.03 | 125.13 | 126.63 | 127.70 | 123.91 | 128.03 |
| Sum(HV) | 12.71 | 12.99 | 15.69 | 15.83 | 15.92 | 17.46 | 18.09 | 15.79 | 18.42 |
| | | | | | | | | | |
| Total $\alpha_{zz}$ | | | | | | | | | |
| CCSD | 132.93 | 133.47 | 140.59 | 139.86 | 141.04 | 144.08 | 145.79 | 139.70 | 146.45 |
| CCSD(T) | 133.65 | 134.25 | 141.68 | 141.10 | 142.20 | 145.50 | 147.35 | 140.94 | 148.08 |

**Table 6** Basis set dependency of the NOs contributions to longitudinal $\gamma_{zzzz}$ of hexatriene. Decomposed values were obtained at the CCSD level of theory. Additionally, CCSD(T) total values are given.

| C:          | VDZ    | VDZ+s  | VDZ+p  | VDZ+d  | VDZ+sp | VDZ+p  | AVDZ   | VDZ    | AVDZ   |
| H:          | VDZ    | VDZ    | VDZ    | VDZ    | VDZ    | VDZ+p  | VDZ    | AVDZ   | AVDZ   |
|-------------|--------|--------|--------|--------|--------|--------|--------|--------|--------|
| $\sigma$(FV)  | -11175 | -11591 | -14565 | -12884 | -14905 | -15929 | -15725 | -14256 | -16120 |
| $\sigma$(HV)  | 4006   | 5105   | 6368   | 5196   | 7349   | 6941   | 7352   | 7887   | 8057   |
| $\sigma$(all) | -7169  | -6486  | -8198  | -7688  | -7557  | -8988  | -8373  | -6369  | -8063  |
| $\pi$(FV)     | 109754 | 110375 | 130891 | 111891 | 131418 | 131879 | 128768 | 119306 | 129463 |
| $\pi$(HV)     | 5756   | 5779   | 34215  | 11305  | 34681  | 39289  | 37697  | 16528  | 39135  |
| $\pi$(all)    | 115511 | 116154 | 165106 | 123195 | 166099 | 171169 | 166465 | 135834 | 168598 |
| Sum(FV)     | 98579  | 98784  | 116326 | 99006  | 116513 | 115950 | 113043 | 105050 | 113343 |
| Sum(HV)     | 9762   | 10883  | 40583  | 16501  | 42029  | 46231  | 45049  | 24415  | 47192  |
| Total $\gamma_{zzzz}$ | | | | | | | | | |
| CCSD        | 108342 | 109668 | 156909 | 115507 | 158542 | 162181 | 158092 | 129465 | 160535 |
| CCSD(T)     | 104299 | 105860 | 155439 | 112562 | 157482 | 161993 | 158102 | 126938 | 161292 |